\documentclass[namedreferences]{solarphysics}
\usepackage[optionalrh,solaromanenum]{spr-sola-addons} 
\usepackage{graphicx}        
\usepackage{fixltx2e}
\usepackage{bm}

\usepackage[update,prepend]{epstopdf}
\usepackage[british]{babel}
\usepackage{color}           
\usepackage{url}             
\usepackage{gensymb, amssymb, textcomp}
\usepackage{hyperref}
\usepackage{float}

\newcommand{\aap}{{\it Astron. Astrophys.}}

\newcommand{\apjl}{{\it Astrophys. J. Lett.}}

\newcommand{\apjs}{{\it Astrophys. J. Supp.}}

\newcommand{\stereo}{STEREO-A/COR1}
\newcommand{\lasco}{SOHO/LASCO-C2}
\begin{document}

\begin{article}

\begin{opening}

\title{Strength of the Solar Coronal Magnetic field - A Comparison of  
Independent Estimates Using Contemporaneous Radio and Whitelight Observations}

\author[addressref=aff1,corref,email={anshu@iiap.res.in}]{\inits{Anshu}\fnm{Anshu}~\lnm{Kumari}}
\author[addressref=aff1,corref,email={ramesh@iiap.res.in}]{\inits{R.}\fnm{R.}~\lnm{Ramesh}}
\author[addressref=aff1,corref,email={kathir@iiap.res.in}]{\inits{C.}\fnm{C.}~\lnm{Kathiravan}}
\author[addressref=aff2,corref,email={tongjiang.wang@nasa.gov}]{\inits{T. J.}\fnm{T. J.}~\lnm{Wang}}

\address[id=aff1]{Indian Institute of Astrophysics, 2nd Block, Koramangala, Bangalore-560034, India}
\address[id=aff2]{Department of Physics, The Catholic University of America and 
                 NASA Goddard Space Flight Center, Code 671, Greenbelt, MD 20771, USA}
\runningauthor{Anshu \textit{et al.}}
\runningtitle{Solar Coronal Magnetic field}

\begin{abstract}

We estimated the coronal magnetic field strength ($B$) during the 23 July 2016 
coronal mass ejection (CME) event using i) the flux rope structure of the CME in the 
whitelight coronagraph images and ii) the band splitting in the associated type {\sc II} burst. 
No models were assumed for the coronal electron density ($N(r)$) used in the estimation.
The results obtained using the above two independent methods correspond to different heliocentric distances ($r$) in the range $\approx$\,2.5\,--\,4.5$\rm R_{\odot}$, but they show excellent consistency and could be fitted with a single power-law distribution of the type $B(r)=5.7r^{-2.6}$ \textrm{G}, which is applicable in the aforementioned distance range. The power law index (\textit{i.e.} $-2.6$) is in good agreement with the results obtained in previous studies by different methods.   

\end{abstract}

\end{opening}

\section{Introduction}
\label{S-Introduction} 

Estimation of the magnetic field strength ($B$) in the solar corona is one of the widely pursued areas of research in observational solar physics. Circular polarization observations in the radio frequency range 1\,--\,20 GHz have been extensively used to measure $B$ above active regions. The corresponding heliocentric distance range is typically $r$\,$\approx$\,1.05\,--\,1.10$\rm R_{\odot}$ (see for example \opencite{Gelfreikh2004}; \opencite{Ryabov2004}; \opencite{White2004}; and the references therein).
Observations of Stokes \textit{V} profiles of the coronal emission line Fe {\sc XIII} {$\lambda$10747} resulting from the longitudinal Zeeman effect were used by \inlinecite{Lin2000} to estimate $B$ over $r$\,$\approx$\,1.12\,--\,1.15$\rm R_{\odot}$. The above measurements in the radio and optical/infrared regimes are limited to the inner corona ($r \rm \lesssim 1.2R_{\odot}$). Moving over to the outer corona ($r > \rm 3R_{\odot}$), Faraday rotation observations 
are generally used to derive the magnetic field \citep{Patzold1987,Spangler2005,Mancuso2013}.
Of late, the standoff distance of a CME
driven shock has also been used to estimate $B$ in the range $r$\,$\approx$\,3\,--\,15$\rm R_{\odot}$ \citep{Kim2012}, and $r$\,$\approx$\,60\,--\,215$\rm R_{\odot}$ \citep{Poomvises2012}. Compared to the above two regions of the corona, estimates of $B$ in the middle corona ($r$\,$\approx$\,1.2\,--\,3.0$\rm R_{\odot}$) are primarily using the different types of transient non-thermal 
radio emission observed at low frequencies ($\lesssim$ 150 MHz) with ground based instruments.
The radio methods, though indirect in contrast to the Zeeman effect in optical wavelengths, 
have provided the bulk of the quantitative information on $B$ in the above distance range
(see \opencite{Dulk1978} for a review on the topic).
The various related techniques employed presently are based on band splitting in type {\sc II} bursts, second harmonic plasma emission in type {\sc II}, type {\sc III}, and type {\sc IV} bursts, gyrosynchrotron emission in type {\sc IV} bursts, Alf\'{v}en speed in type {\sc II} bursts, quasi-periodicity in type {\sc III} bursts, \textit{etc}. Note that while the mechanism is either fundamental  and/or second harmonic plasma emission in the case of type {\sc II} and type {\sc III} bursts, it is either second harmonic plasma or gyrosynchrotron emission in the case of type {\sc IV} bursts. We would like to point out here that characteristics of the polarization of fundamental plasma emission from solar radio bursts is yet to be established. So, estimates of $B$ are currently possible only using second harmonic plasma emission from the bursts. The estimates are from observations of either the total intensity, \textit{i.e}. 
Stokes \textit{I}, alone
(\opencite{Smerd1975}; \opencite{Gopalswamy1990}; \opencite{Bastian2001};
\opencite{Vrsnak2002}; \opencite{Mancuso2003}; \opencite{Ramesh2003}; \citeyear{Ramesh2004};
\citeyear{Ramesh2013}; \opencite{Cho2007}; \opencite{Kishore2016}) or
both the total and circularly polarized intensities, \textit{i.e}. Stokes \textit{I} and \textit{V}
(\opencite{Dulk1980}; \opencite{Gary1985}; \opencite{Ramesh2010b}; \opencite{Ramesh2011b}; \opencite{Tun2013}; 
\opencite{Sasikumar2013}; \citeyear{Sasikumar2014};
\opencite{Hariharan2014}; \citeyear{Hariharan2016a}; \opencite{Anshu2017}). 
Note that we have mentioned only Stokes \textit{I} and \textit{V} emission here since differential Faraday rotation 
of the plane of polarization in the solar corona and Earth's ionosphere makes it impossible to observe the 
linear polarization (represented by Stokes \textit{Q} and \textit{U}) within the typical observing bandwidths of $\approx$\,100 kHz (see for example \opencite{Grognard1973}.)
It has been shown that thermal radio emission observed from the solar corona at low frequencies can also be used to estimate $B$ \citep{Ramesh_apj2003,Sastry2009,Ramesh2010}.

\inlinecite{Gopalswamy2012} reported a new technique, based on the geometrical properties
of the flux rope structure of a whitelight CME and the associated shock speed,
to estimate $B$. Similarly, \citeauthor{Kwon2013a} (\citeyear{Kwon2013a}; \citeyear{Kwon2013b}) reported estimates of $B$ from measurements of the propagation speed of a fast magnetosonic wave associated with a CME. But, 
reports of comparison between 
the $B$ values obtained independently in radio and optical wavelengths for the same event using simultaneous observations which do not assume any model for $N(r)$ are rare. 
The comparison is important to validate the different techniques to estimate $B$. 
Considering this, in the present work we take advantage of the simultaneous observations of a whitelight CME near the Sun with flux rope structure and band splitting exhibited by the associated type {\sc II} burst, we estimate the magnetic field strength in the middle corona.

\section{Observations}

\subsection{Radio Observations}
\label{S2.1}

The radio data were obtained with the \textit{Gauribidanur RAdio SpectroPolarimeter} (GRASP; \opencite{Kishore2015},
\opencite{Hariharan2016b}) with a new digital back-end correlator receiver, operated at the Gauribidanur radio observatory\footnote{See http://www.iiap.res.in/centers/radio.} about 100 km north of Bangalore
in India \citep{Ramesh2011a}. 
The primary receiving elements of GRASP are two log-periodic dipole antennas (LPDAs) oriented orthogonal to each other. The characteristics of these LPDAs are: half-power beam width (HPBW) $\approx$\,$100^\circ$ (in both the 
E- and H-planes), gain G $\approx$\,5.5 dBi\footnote{The gain of an antenna system relative to an isotropic radiator.} and effective collecting area $\rm A_{e} \approx$\,0.3$\lambda^2$.
The radio frequency (RF) signals from the two LPDAs are transmitted to the receiver room using independent low loss coaxial cables where a 2-channel, 8-bit Analog-to-Digital Converter (ADC) is used to directly digitize the RF signal from each LPDA. The maximum possible sampling clock for the ADC is $\approx$\,250 MHz. Since the RF signals received by the LPDAs in the GRASP are limited to $\lesssim$ 100 MHz, they were sampled at 200 MHz.
The ADC was interfaced to a ROACH\footnote{See https://casper.berkeley.edu/.} board.
The signals from the two channels of the ADC ($E_x$ and $E_y$, respectively) were Fourier transformed to obtain the instantaneous spectrum with a spectral resolution of $\approx$\,200 kHz and temporal resolution of $\approx 100$ msec. The spectrum thus obtained is used to compute Stokes \textit{I} and \textit{V} intensities as follows \citep{Collett1992}:

\begin{equation}
I=<E_xE_x^* + E_yE_y^*>
\end{equation}
\begin{equation}
V=-2i <E_xE_y^* - E_yE_x^*>
\end{equation}

\begin{figure}
\centering
\centerline{\includegraphics[angle=0,height=10cm,width=11cm]{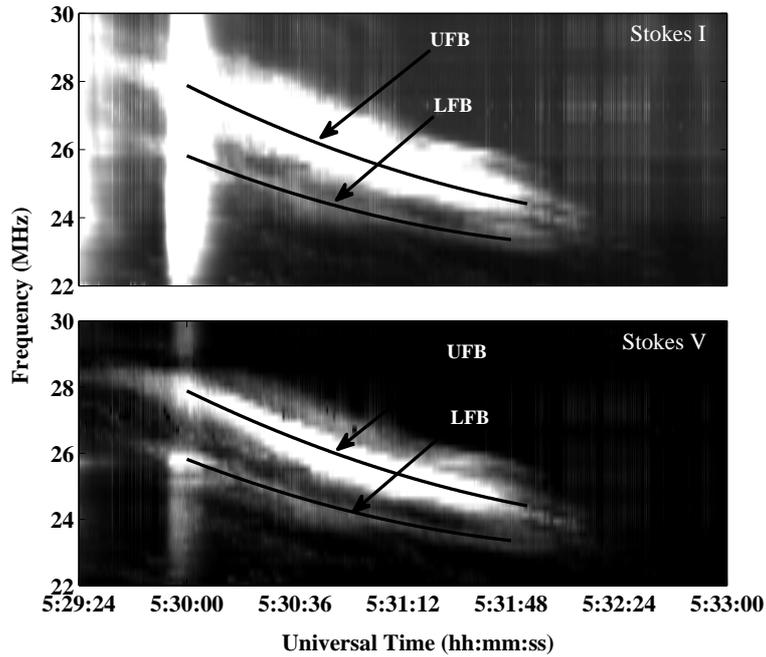}}
\caption{Type {\sc V} burst ($\approx$\,05:30 UT) 
and type {\sc II} radio burst ($\approx$\,05:30\,--\,05:32 UT) observed with the GRASP
on 23 July 2016. The band-splitting of the type {\sc II} burst can be clearly noticed in both 
the Stokes \textit{I} and \textit{V} spectra. The \textit{black} lines indicate the upper and lower frequency bands of the split-band emission. The thin, closely spaced vertical lines seen in the spectra are artefacts.}
\label{fig:figure1}
\end{figure}

Figure \ref{fig:figure1} shows the dynamic spectrum of the transient emission from the Sun observed with the GRASP on 23 July 2016. The fast drifting feature (from high to low frequencies) at $\approx$\,05:30 UT is a type {\sc V} burst. The slow drifting feature between $\approx$\,05:30\,--\,05:32 UT is a type {\sc II} burst.
While the type {\sc II} burst is confined to a narrow range of frequencies (start frequency $\approx$\,28 MHz; end frequency $\approx$\,24 MHz), the type {\sc V} burst extends beyond. We verified the aforementioned frequency limits from observations with the 
\textit{Gauribidanur LOw-frequency Solar Spectrograph} (GLOSS; \opencite{Ebenezer2001}; 
\citeyear{Ebenezer2007}; \opencite{Kishore2014})
and the e-CALLISTO (\opencite{Monstein2007}; \opencite{Benz2009}) in the frequency range $\approx$\,40\,--\,450 MHz. 
Note that observations of type {\sc II} bursts over a limited spectral range as in the present work are not uncommon (see for example \opencite{Hariharan2015}). We will discuss the above type {\sc II} burst in the rest of this work. 

Solar type {\sc II} bursts are signatures of magnetohydrodynamic (MHD) shock waves travelling outwards in the corona. These shocks accelerate electrons to suprathermal velocities. The energetic electrons excite Langmuir (plasma) waves which are converted into radio waves escaping the corona (see
\opencite{Roberts1959}; \opencite{Wild1963}; \opencite{Nelson1985}; \opencite{Mann1995};
\opencite{Aurass1997}; \opencite{Gopalswamy2006} for details).
A closer inspection of the type {\sc II} burst in Figure \ref{fig:figure1} indicates that it is split, and 
appears as two bands (upper frequency band [UFB] and lower frequency band [LFB]). The frequency ranges are
$\approx$\,28.0\,--\,24.2 MHz (UFB) and $\approx$\,26.0\,--\,23.5 MHz (LFB). 
The LFB and UFB bands relate to emission from the coronal regions ahead of and behind the associated
MHD shock, respectively \citep{Vrsnak2001}.

\subsection{STEREO-A/COR1 and SOHO/LASCO-C2 Observations}
\label{S2.2}

\begin{figure}
\centering
\centerline{\includegraphics[angle=0,height=7cm,width=16cm]{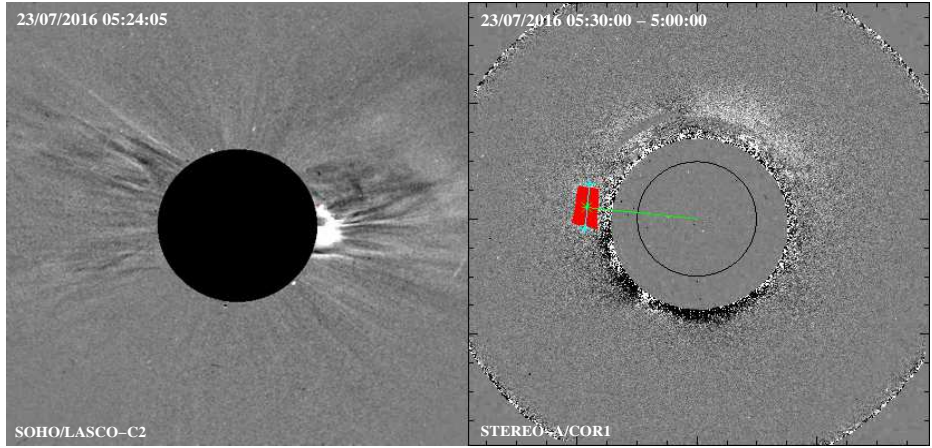}}
\caption{\textit{Left Panel:} SOHO/LASCO-C2 coronagraph difference image of the CME that occurred on 23 July 2016 around $\approx$\,05:24 UT. 
The \textit{black} circle (radius $\approx$\,$2.2 \rm R_{\odot}$) represents the occulting disk of the coronagraph. Solar north 
is straight up and east is to the left in the image.
\textit{Right Panel:} STEREO-A/COR1 pB difference image of the same CME as in the left panel. The \textit{black} and the
\textit{gray} circles indicate the solar limb (radius\,=\,$\rm 1R_{\odot}$) and the coronagraph 
occulter (radius\,$\approx$\,$1.4 \rm R_{\odot}$), respectively. The CME is fainter compared to the 
SOHO/LASCO-C2 observations
since its angle with respect to the plane-of-sky (POS) is relatively larger for STEREO-A (see Section \ref{S2.2} for details).
The electron density of the CME mentioned in Section 3.1 was estimated from the coronal region covered by the
\textit{red} rectangular box. 
The vertical line across the above box represents the width of the CME 
($\approx$\,$\rm 0.79R_{\odot}$).}
\label{fig:figure2}
\end{figure}

The optical data were obtained with  the \textit{Large Angle and Spectrometric Coronagraph C2} (LASCO-C2; \opencite{Brueckner1995})
on board the \textit{Solar and Heliospheric Observatory} (SOHO), and the COR1 coronagraph of the \textit{Sun-Earth Connection Coronal and Heliospheric Investigation} (SEECHI; \opencite{Howard2008}) on board the \textit{Solar Terrestrial Relationship Observatory} (STEREO).
The type {\sc II} burst in Figure \ref{fig:figure1} was associated with a whitelight CME observed by the SOHO/LASCO-C2 and STEREO-A/COR1 coronagraphs. The CME was noticed in the SOHO/LASCO-C2 field-of-view (FOV) around $\approx$\,05:24 UT (see 
left panel in Figure \ref{fig:figure2}).
Its central position angle (PA, measured counter-clockwise from the solar north) 
during the above epoch is $\approx$\,$267^{\circ}$, and the angular width is 
$\approx$\,$65^{\circ}$.  
The projected heliocentric distance of the CME is $\approx$\,$\rm 2.79R_{\odot}$\footnote{See http://spaceweather.gmu.edu/seeds/monthly.php?a=2016\&b=07.}.
There was a 3B class H$\alpha$ 
flare from AR12565 located on the solar disk at heliographic coordinates 
N02W75. The flare was observed during the interval $\approx$\,05:09\,--\,06:33 UT. The peak was at $\approx$\,05:31 UT\footnote{See ftp://ftp.ngdc.noaa.gov/STP/swpc\_products/daily\_reports/solar\_event\_reports/2016/07.}.
The X-Ray Sensor (XRS) on board the \textit{Geostationary Operational Environmental Satellite} (GOES-15) observed a 
M5.5 class soft X-ray flare from the above active region, during the 
interval $\approx$\,05:27\,--\,05:33 UT. The peak time was the same as that of the H$\alpha$ flare, \textit{i.e}. $\approx$\,05:31 UT. This indicates that the onset of the type {\sc II} burst ($\approx$\,05:30 UT) in Figure \ref{fig:figure1} was close to the peak time
of the flare.

The right panel in Figure \ref{fig:figure2} shows the
STEREO-A/COR1 difference image of the solar corona obtained on 23 July 2016 at 
$\approx$\,05:30 UT. 
The boxed region at PA $\approx$\,$84^{\circ}$ indicates the same CME noticed in the SOHO/LASCO-C2 
observations in the left panel of Figure \ref{fig:figure2} at $\approx$\,5:24 UT.
Its projected heliocentric distance in the STEREO-A/COR1 FOV is $\approx$\,$\rm 1.86R_{\odot}$. 
The subtracted reference image for STEREO-A/COR1 in Figure \ref{fig:figure2}  was observed at $\approx$\,05:00 UT. 
We found that STEREO-A was at $\approx$\,E$153^{\circ}$ during the
above observations\footnote{See stereo-ssc.nascom.nasa.gov/cgi-bin/make{\_}where.gif.}. 
Therefore for the STEREO-A view, the flaring region is on the far side of the Sun 
at $\approx$\,$42^{\circ}$ behind the east limb.
For the SOHO/LASCO-C2 view, the above region is located on the solar disk at 
$\approx$\,$15^{\circ}$ from the west limb.
The CME observations will have projection effects in view of the above disk locations. 
We calculated the de-projected heliocentric distances of the CME 
from the SOHO/LASCO-C2 and STEREO-A/COR1 images assuming that the projection effects vary as 1/cos($\phi$), 
where $\phi$ is the angle from the plane-of-sky (POS). 

\section{Analysis and Results}
\label{S-Analysis and Results}  

\subsection{Estimates of $N(r)$ Using Polarized Brightness (pB) 
Measurements with STEREO-A/COR1 and SOHO/LASCO-C2}
\label{pB measurements}

The electron density in the corona depends upon the brightness and the polarization in the corona \citep{Van1950}.
We used polarized Brightness (pB) measurements with the STEREO-A/COR1 and SOHO/LASCO-C2 coronagraphs (see Figure \ref{fig:figure2}) to find a suitable model for $N(r)$ to estimate the $B(r)$ values.  
In the case of STEREO-A/COR1, the $N(r)$ of the background corona 
and the CME were calculated using the images obtained at $\approx$\,05:00 UT 
(\textit{i.e.} before the onset of the CME) and $\approx$\,05:30 UT (during the type II burst), respectively.
These calculations were done using the spherically symmetric inversion technique \citep{Wang2014}.
The $N(r)$ of the background corona was calculated in the distance range 
$r \rm \approx$\,$\rm 1.5-3.7R_{\odot}$ and over PA $\approx$\,$84^{\circ}$\,$\pm 5^{\circ}$
(see right panel in Figure \ref{fig:figure2}). In case of the CME, the 
background subtracted pB radiation from the boxed region in the right panel of Figure \ref{fig:figure2} was 
used. 
The de-projected heliocentric distance of the latter is $\approx$\,$\rm 2.63R_{\odot}$.
We assumed that the line-of-sight (LOS) depth of the CME is equal to its width, and the CME is at an angle of $\approx$\,$42^{\circ}$ relative to the POS (see Section \ref{S2.2}) for the above calculations. The $N(r)$ of the background corona and CME 
is $\rm \approx$\,$0.37 \times \rm 10^{6}~cm^{-3}$ and $\rm \approx$\,$\rm (1.8 \pm 3.3) \times 10^{6}~cm^{-3}$, respectively at $r \approx$\,$\rm 2.63R_{\odot}$.

\begin{figure}
\centering
\centerline{\includegraphics[angle=0,height=10cm,width=11cm]{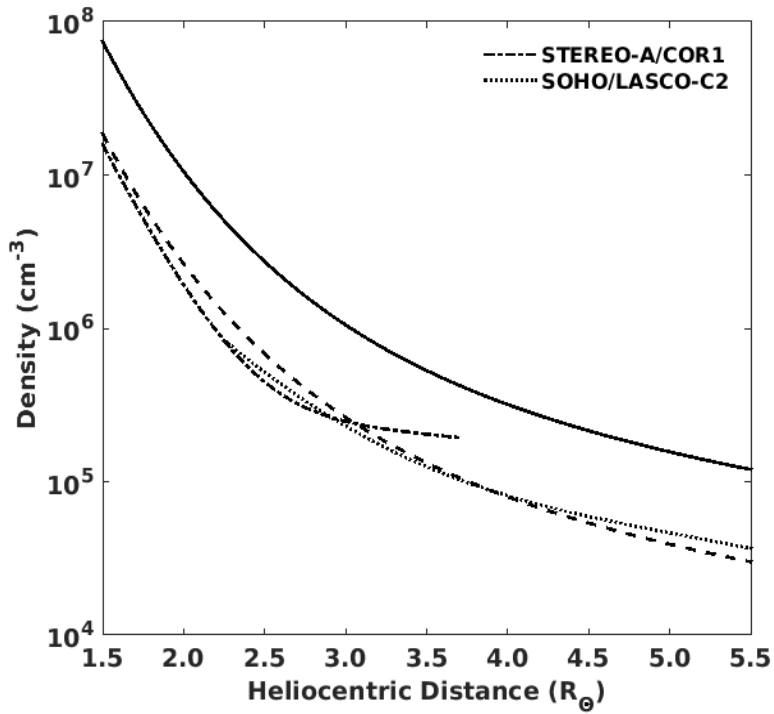}}
\caption{Estimates of $N(r)$ in the background corona using pB measurements 
with STEREO-A/COR1 ($r \rm \approx$\,1.5\,--\,3.7$\rm R_{\odot}$) and SOHO/LASCO-C2
 ($r \rm \approx$\,2.3\,--\,5.5$\rm R_{\odot}$) 
corongraphs. The \textit{dashed} line ($r \rm \approx$\,$1.5$\,--\,$\rm 5.5R_{\odot}$)
is the best fit to the measurements of $N(r)$ with the above two instruments. The \textit{solid} line in the same
distance range represents $4 \times$ the density values corresponding to the above fit.}
\label{fig:figure3}
\end{figure}

In the case of SOHO/LASCO-C2, $N(r)$ of the background corona was calculated 
over the range PA $\approx$\,$256^{\circ} \pm 5^{\circ}$ using the image obtained at $\approx$\,02:54 UT (see left panel in Figure \ref{fig:figure2}). The distance range was 
$r \approx$\,2.3\,--\,5.5$\rm R_{\odot}$. 
At the de-projected heliocentric distance $\approx$\,2.63$\rm R_{\odot}$, the background density is $\rm \approx$\,$0.43 \rm \times 10^{6}~cm^{-3}$.
Considering this along with the previously mentioned similar measurements with STEREO-A/COR1,  the average $N(r)$ value in the background corona at the location of the CME ($r \approx \rm 2.63~R_{\odot}$) is $\approx$\,$0.4 \rm \times 10^{6}~cm^{-3}$ (see right panel in Figure \ref{fig:figure2}). 
This implies the total $N(r)$ at the CME location is $\rm \approx$\,2.2 $\times \rm 10^{6}~cm^{-3}$ (CME \textit{plus} background corona).


Figure \ref{fig:figure3} shows the background density estimates for STEREO-A/COR1 and SOHO/LASCO-C2 for the distance range 
$r \rm \approx$\,1.5\,--\,3.7$\rm R_{\odot}$ and $r \approx$\,2.3\,--\,6.4$\rm R_{\odot}$, respectively. 
We found that the best fit to the background density is 
$N_{\textsubscript{cor}}(r) = 1.521 \times 10^{8} r^{-7.279} + 1.84 \times 10^{8} r^{-7.938} + 
2.07 \times 10^{7} r^{-4.852} + 7.52 \times 10^{5} r^{-2.024}$, where the first two terms correspond
to $r \rm \approx$\,1.5\,--\,2.5$\rm R_{\odot}$ and $r \rm \approx$\,2.5\,--\,3.7$\rm R_{\odot}$ in the STEREO-A/COR1 FOV,
and the last two terms correspond to $r \rm \approx$\,2.3\,--\,4.0$\rm R_{\odot}$ and $r \rm \approx$\,4.0\,--\,5.5$\rm R_{\odot}$ 
in the SOHO/LASCO-C2 FOV, respectively. Note that 
the widely used Saito model for $N(r)$, derived from similar corongraph
observations in the range $r \rm \approx$\,2.5\,--\,5.5$\rm R_{\odot}$ \citep{Saito1977}, has also  
two different radial gradients as in the present case.


The plasma frequency ($f_{\textsubscript{p}}$) corresponding to the estimates of the 
total density ($\rm \approx 2.2 \times 10^{6}~cm^{-3}$) 
is $\approx$ 13.3 MHz. This implies that the type {\sc II} burst in Figure \ref{fig:figure1} is most likely due to second harmonic 
($2f_{\textsubscript{p}}$) emission since its mean frequency at $\approx$ 05:30 UT is $\approx$ 27 MHz. The estimated maximum degree of circular polarization dcp is $\approx$\,18 \% and average dcp is $\approx$\,14 \% of the type {\sc II} burst, which are also consistent with that expected for harmonic plasma emission \citep{Dulk1980}.
We would like to point out here that there are reports of third harmonic emission ($3f_{\textsubscript{p}}$) of type {\sc II} bursts
also in the literature. But such observations are rare (\opencite{Roberts1959}; \opencite{Bakunin1990}:
\opencite{Kliem1992}; \opencite{Zlotnik1998}).
Interestingly, GLOSS observed the third harmonic emission of the above type {\sc II} event in the frequency range $\approx$\,56\,--\,43 MHz\footnote{http://www.iiap.res.in/files/solarradioimages/gbd/GLOSS\_20160723.jpg.}.
This was possible most likely because of the comparatively larger collecting area of GLOSS.
Note that the latter consists of eight antennas for radio frequency signal reception as compared to the typical solar radio
spectrographs where only a single antenna is used (see for example \opencite{Benz2009}). 
Further, Sun was closer to the local zenith
in Gauribidanur during the observations. The antenna gain is always maximum in that direction of the sky.

\subsection{Speed of the CME}

\begin{figure}
\centering
\centerline{\includegraphics[angle=0,height=9cm,width=11cm]{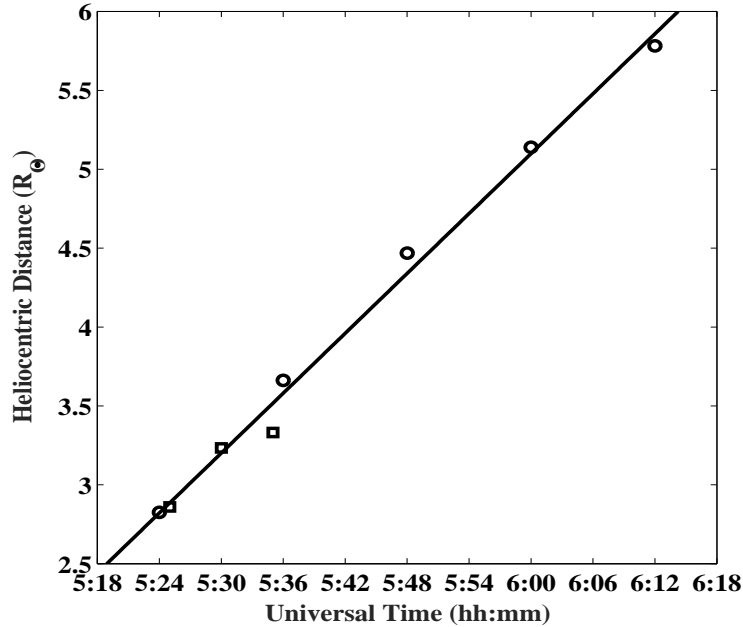}}
\caption{Height-time \textit{(h-t)} plot showing the de-projected heliocentric distances of the CME LE
observed with the STEREO-A/COR1 coronagraph (indicated by \textit{squares}) and SOHO/LASCO-C2 coronagraph (indicated by \textit{circles}) on 
23 July 2016 at different epochs. The \textit{solid} line is the linear fit to the estimates.}
\label{fig:figure4}
\end{figure}

The CME was noticed in the STEREO-A/COR1 FOV during $\approx$\,05:25\,--\,05:35 UT. In the case of SOHO/LASCO-C2, 
it was in the interval $\approx$\,05:24\,--\,06:12 UT.
The deprojected height-time \textit{(h-t)} plot of the CME LE obtained from the above observations is shown in Figure \ref{fig:figure4}. The simultaneous measurements of the CME LE location with the above two instruments 
during $\approx$\,05:24\,--\,05:36 UT are reasonably consistent.
This confirms that the SOHO/LASCO-C2 and STEREO-A/COR1 observations described above correspond to the same CME. 
Its estimated speed ($v_{\textsubscript{CME}}$) based on the linear-fit to the \textit{(h-t)} measurements in Figure \ref{fig:figure4} 
is $\approx$\,735 km/s.

\subsection{Estimates of $B(r)$}

\subsubsection{\textbf{Radio Observations}}\label{s3.3.1}

The pB measurements using STEREO-A/COR1 and SOHO/LASCO-C2 coronagraphs described in Section \ref{pB measurements} 
indicate that $N(r) \rm \approx$\,2.2 $\times \rm 10^{6}~cm^{-3}$ (CME \textit{plus} background corona)
at $r \approx$\,2.63$\rm R_{\odot}$. This is $\approx$\,$4N_{\textsubscript{cor}}(r)$ of the background corona at the 
above location (see Figure \ref{fig:figure3}).
We used the UFB and LFB bands of the type {\sc II} burst (marked with black lines in Figure \ref{fig:figure1}) 
to calculate $B(r)$ using the split-band technique (see for example \opencite{Vrsnak2002}; \opencite{Cho2007}). The drift rate of the burst is $\approx$\,0.35 MHz/s. 
We converted this drift rate of the burst to shock speed ($v_{\textsubscript{S}}$) using the 
4$N_{\textsubscript{cor}}(r)$ model mentioned earlier (see Section 3.1). 
The mean value of $v_{\textsubscript{S}}$ for the four different spectra (\textit{i.e}. LFB and UFB in Stokes \textit{I} and \textit{V}) 
is $\approx$ 800 km/s. This is nearly same as the $v_{\textsubscript{CME}}$     
in Figure \ref{fig:figure4}, implying the type {\sc II} burst is most likely due to the MHD shock driven by the 
CME LE (\opencite{Ramesh2010a}; \opencite{Ramesh2012}). The low start frequency of the type {\sc II} burst also 
indicates the same \citep{Gopalswamy2006}.
We then calculated $B(r)$ using the following equation:
\begin{equation}
B (G) = 5.1\times10^{-5}v_{\textsubscript{A}}f_{\textsubscript{p}}
\end{equation}
where $v_{\textsubscript{A}}$ is the Alf\'{v}en speed. It was obtained using the parameters like 
instantaneous bandwidth,
density jump, Alf\'{v}enic Mach number $M_{\textsubscript{A}}$
related to the split-band type burst in Figure \ref{fig:figure1} (see \opencite{Vrsnak2002}
for the relationship between the above parameters), and the 
relationship, $v_{\textsubscript{A}} = v_{\textsubscript{S}}/M_{\textsubscript{A}}$.

%
The results indicate that $B(r) \approx$\,(0.47\,--\,0.44) $\pm$ 0.02 G in the range
$r \rm \approx$\,2.61\,--\,2.74$\rm R_{\odot}$.
Note that the above distance interval corresponds to $f_{\textsubscript{p}} \approx$\,14\,--\,12 MHz (the fundamental component of the split-band type {\sc II} burst in the present case, see Figure \ref{fig:figure1} and Section 2.2) since the 
$2f_{\textsubscript{p}}$ emission occurring at the same time 
as the $f_{\textsubscript{p}}$ emission should be generated at the same location as the 
$f_{\textsubscript{p}}$ emission \citep{Smerd1962}. 

\subsubsection{\textbf{Whitelight Observations}}

\begin{figure}
\centering
\centerline{\includegraphics[angle=0,height=7cm,width=16cm]{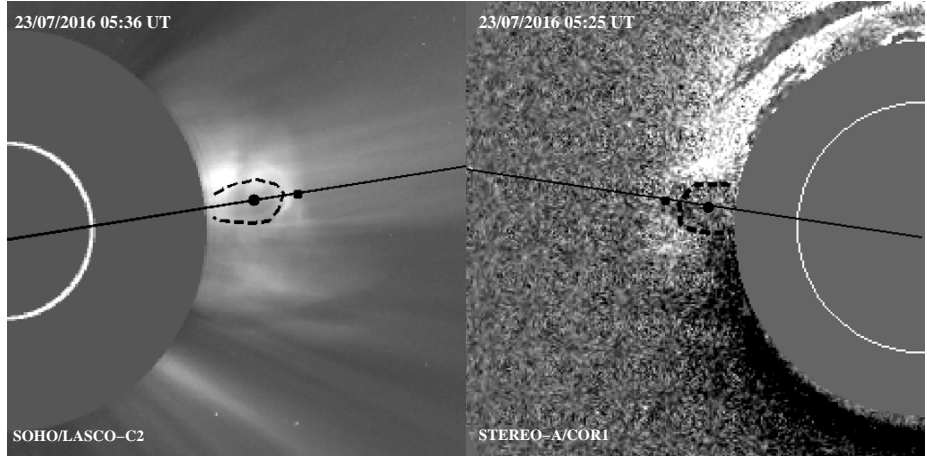}}
\caption{\textit{Left Panel:} Same as the observations in the left panel of Figure \ref{fig:figure2}, 
but obtained at $\approx$\,05:36 UT. The \textit{white} circle indicates the solar limb (radius =\,$\rm 1R_{\odot}$). The different features of the CME used for the estimates of $B$ (see Table \ref{tab:table1}) are:
\textit{black square} -- location of the CME shock, \textit{black dashed} lines - CME flux rope,
\textit{black dot} - center of the circle fitted to the flux rope.  \textit{Right Panel:} Same as the observations in the right panel 
of Figure \ref{fig:figure2}, but obtained at $\approx$\,05:25 UT. The \textit{white} circle indicates the solar 
limb (radius =\,$\rm 1R_{\odot}$). 
The different features of the CME used for the estimates of $B$ (see Table \ref{tab:table1}) are:
\textit{black square} -- location of the CME shock, \textit{black solid} lines -- CME flux rope,
\textit{black dot} -- center of the circle fitted to the flux rope (see Gopalswamy et al., 2012 for details). 
The \textit{black solid} line connects the aforementioned CME features with the center of the Sun.}
\label{fig:figure5}
\end{figure}

We also estimated $B(r)$ using the whitelight images obtained with STEREO-A/COR1 (right panel in 
Figure \ref{fig:figure5}) 
and SOHO/LASCO-C2 (left panel in \ref{fig:figure5}), and the shock standoff distance technique described 
in \citet{Gopalswamy2012}. The basic CME related parameters used in the above technique, i.e. 
the heliocentric distance of the CME flux rope $R_{\textsubscript{fl}}$, leading edge of the shock 
$R_{\textsubscript{sh}}$ and center of curvature of the flux rope $R_{\textsubscript{c}}$
were measured for the aforesaid images as well as their next images observed with the respective coronagraphs. 
We assumed the adiabatic exponent $\gamma$ to be $4/3$ for the calculations.
The stand-off distance $\Delta R$, $M_{\textsubscript{A}}$ and $v_{\textsubscript{s}}$ were calculated 
using the above parameters (see \cite{Gopalswamy2012}
for the relationship between the above parameters), and $v_{\textsubscript{A}}$ was 
estimated as earlier (Section \ref{s3.3.1}). Adopting the same 4$N_{\textsubscript{cor}}(r)$ model as in the
$B(r)$ estimate using the radio observations, we estimated the 
density for {\stereo} and {\lasco} at $r \rm \approx$\,3.11$\rm R_{\odot}$ and $r \rm \approx$\,4.40$\rm R_{\odot}$, respectively.
The magnetic field strength in the corona was estimated using Equation 3,
and the values are $B(r) \approx$\,0.30 G at $r \approx$\,$\rm 3.11R_{\odot}$ (STEREO-A/COR1) and
$B(r) \approx$\,0.12 G at $r \approx$\,$\rm 4.40R_{\odot}$ (SOHO/LASCO-C2). 
The parameters related to the above calculations are listed in Table \ref{tab:table1}.


\begin{table}
\caption{Parameters related to the estimates of $B(r)$ from whitelight observations.}
\label{tab:table1}
\begin{tabular}{ccccccc}     
\noalign{\smallskip}\hline\noalign{\smallskip}
\textrm{Time} & $R_{\textsubscript{sh}}$\tabnote{Heliocentric distance of the CME shock.} &  
$R_{\textsubscript{fl}}$\tabnote{Heliocentric distance of the CME flux rope LE.} &  
$R_{\textsubscript{c}}$\tabnote{Radius of curvature of the circle fitted to the flux rope in Figure 5.} 
& $\Delta{R}$ & 
$N_{\textsubscript{e}}$ & $B$ \\
 & & & & [$R_{\rm sh}$\,--\,$R_{\rm fl}$] & [4\,$\times N_{\rm cor}(r)$] & \\
\textrm{(UT)} & ($\rm R_{\odot}$) & ($\rm R_{\odot}$) & ($\rm R_{\odot}$) & ($\rm R_{\odot}$) & ($\rm \times 10^{5}~cm^{-3}$) & \textrm{(G)}\\
\hline
\textrm{STEREO-A/COR1} \\
05:25 & 2.75 & 2.30 & 0.26 & 0.45 & - & - \\
05:30 & 3.11 & 2.50 & 0.37 & 0.61 & 9.00 & 0.30 \\ \hline
\textrm{SOHO/LASCO-C2} \\
05:36 & 3.54 & 3.37 & 0.36 & 0.17 & - & - \\
05:48 & 4.40 & 4.02 & 0.61 & 0.38 & 2.31 & 0.12\\ \hline
\end{tabular}
\end{table}

\subsection{Results}
\label{res}

\begin{figure}
\centering
\centerline{\includegraphics[angle=0,height=10cm,width=11cm]{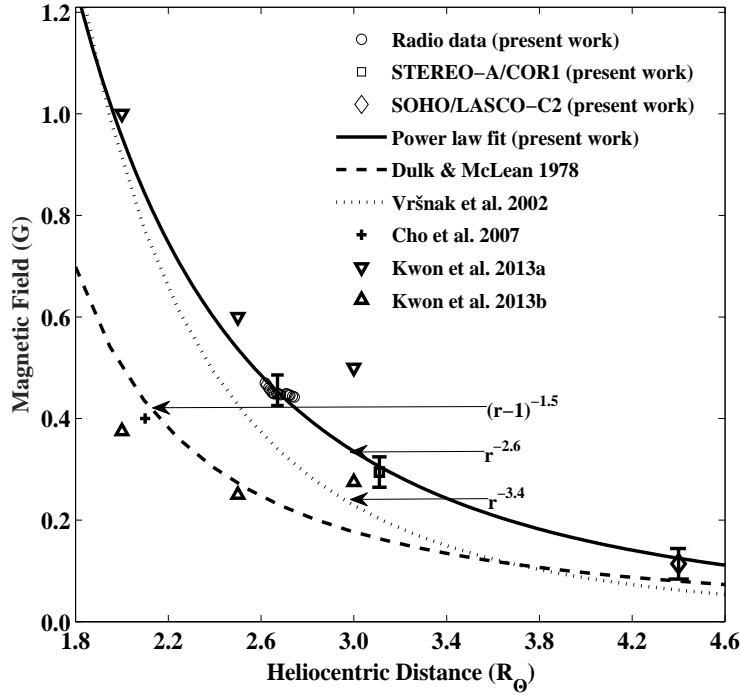}}
\caption{Estimates of $B(r)$ using band-splitting of the type {\sc II} radio burst in Figure \ref{fig:figure1}, and the shock 
stand-off technique applied to the associated whitelight CME observed with the STEREO-A/COR1 and 
SOHO/LASCO-C2 coronagraphs.}
\label{fig:figure6}
\end{figure}

Figure \ref{fig:figure6} shows the combined plot of the different estimates of $B(r)$ mentioned above.
We find that the radial dependence of the data points could be described by a single power-law 
relation of the form $B(r)=5.7r^{-2.6}$ G. 
For comparison, the $B(r)$ models obtained by \inlinecite{Dulk1978}; \inlinecite{Vrsnak2002}, 
and the $B$ values obtained by \citeauthor{Kwon2013a} (\citeyear{Kwon2013b}; \citeyear{Kwon2013a}) 
are also overplotted in Figure \ref{fig:figure6}.
We find that the $B(r)$ curve obtained in this study best matches to the result of \inlinecite{Kwon2013b}.
 


\section{Summary}

We obtained independent estimates of $B(r)$ from the split-band type {\sc II} burst observed on 
23 July 2016, and the associated whitelight CME with flux rope structure. 
Rather than assuming any existing density model for the $N(r)$ values, we used the $N(r)$ values obtained from STEREO-A/COR1 and SOHO/LASCO-C2 pB measurements. 
The $B(r)$ values, spread over the range $r \approx$\,2.5\,--\,4.5$\rm R_{\odot}$, could be fitted by a single power-law 
fit of the form $B(r)=5.7r^{-2.6}$ G. The power-law index (\textit{i.e.}$ -2.6$) is in good agreement with that mentioned in the
literature by different authors for variation of $B(r)$ in the outer corona: i) for example, using
Faraday rotation observations of the linearly polarized carrier signals of the HELIOS spacecraft,
\inlinecite{Patzold1987} found that $B(r)$ varies as $r^{-2.7}$ in the range $r \approx$\,3\,--\,10$\rm R_{\odot}$;
ii) recently, \inlinecite{Mancuso2013} reported  
$B(r)=3.76r^{-2.29}$ G in the range $r \approx$\,5\,--\,14$\rm R_{\odot}$ using observations of similar signals 
from extragalactic radio sources occulted by the solar corona. 
Considering $B(r)$ values at specific $r$ in the outer corona (close to the range in the present work), 
we find that \inlinecite{Spangler2005};
\inlinecite{Gopalswamy2011} had estimated $B(r) \approx$ 39 mG and $\approx$\,48 mG at $r \rm \approx$\,6.2$\rm R_{\odot}$ and
$\rm \approx$\,6.0$\rm R_{\odot}$, respectively. Compared to these, the power-law expression derived in the present work 
predicts $B(r) \approx$\,49 mG and $\approx$\,54 mG at the above distances.
Therefore, unambiguous estimates of $B(r)$ in the middle corona can be obtained using 
radio and whitelight observations as described in the present work upon identifying the appropriate $N(r)$ there.

\begin{acks}

We thank the staff of the Gauribidanur observatory for their
help in maintenance of the antenna receiver  systems and  the  observations.
AK acknowledges Mr. K. Hariharan and Mr. V. Mugundhan for discussions.
We also thank the referee for his/her comments which helped us to present the
results more clearly.
The SOHO data are produced by a consortium of the Naval Research Laboratory (USA),
Max-Planck-Institut fuer Aeronomie (Germany),~Laboratoire
d'Astronomie (France),~and the University of Birmingham (UK). SOHO
is a project of international cooperation between ESA and NASA.
The SOHO-LASCO CME catalog is generated and maintained at the CDAW Data Center
by NASA and the Catholic University of America in cooperation with the
Naval Research Laboratory. The work of TJW was supported by 
NASA Cooperative Agreement NNG11PL10A to CUA.We also thank the referee and guest editors for their comments.\\

\noindent
\textbf{Disclosure of Potential Conflicts of interest}   
We declare that there are no conflicts of interest for the work presented here.  
\end{acks}
%

\bibliographystyle{spr-mp-sola}
\bibliography{reference} 

\end{article}
\end{document}